\documentclass[a4paper,10pt,twoside]{cpc-hepnp}

\usepackage{CJK,upgreek,fancyhdr}
\usepackage{multicol}
\usepackage{graphicx}
\usepackage{booktabs}
\usepackage{amssymb,bm,mathrsfs,bbm,amscd}
\usepackage[tbtags]{amsmath}
\usepackage{lastpage}

\usepackage{natbib}

\usepackage{enumerate}

\usepackage{latexsym}
\usepackage{bm}

\usepackage{hhline,multirow}
\usepackage{dcolumn}
\usepackage{url}
\usepackage[colorlinks=true,linkcolor=blue]{hyperref}
\usepackage{color}
\usepackage{indentfirst}
\usepackage{subfigure}
\usepackage{psfrag}
\usepackage{slashed}
\usepackage{float}
\usepackage{setspace}

\newcommand{\gev}{$\mathrm{GeV}$}

\newcommand{\Tr}{\,{\rm Tr}}

\begin{document}
	

\fancyhead[c]{\small Chinese Physics C~~~Vol. xx, No. x (xxxx) yyyzzz}
\fancyfoot[C]{\small xxxyyy-\thepage}
	
\footnotetext[0]{Received day month 2019}

\title{
	Sea quark contributions to nucleon electromagnetic form factors with the nonlocal chiral effective Lagrangian
	\thanks{
		This work is supported by the National Natural Sciences Foundations of China under the grant No. 11475186,
		the Sino-German CRC 110 ``Symmetries and the Emergence of Structure in QCD" project by NSFC under the grant
		No.11621131001.
	}}

\author{%
	Ming-Yang Yang$^{1,2;1)}$\email{yangmy@ihep.ac.cn}%
\quad Ping Wang$^{1,3;2)}$\email{pwang4@ihep.ac.cn}%
}
\maketitle

\address{
$^1$ Institute of High Energy Physics, CAS, P. O. Box 918(4), Beijing 100049, China \\
$^2$ College of Physics Sciences, University of Chinese Academy of Sciences, Beijing 100049, China \\
$^3$ Theoretical Physics Center for Science Facilities, 
CAS, Beijing 100049, China \\
}

\begin{abstract}
The sea quark contributions to the nucleon electromagnetic form factors from up, down and strange quarks are studied with the nonlocal chiral effective Lagrangian. Both octet and decuplet intermediate states are included in the one loop calculation. Compared with the strange form factors, though their signs are the same, the absolute value of the light quark form factors
are much larger. For both electric and magnetic form factors, the contribution from $d$ quark is larger than that from $u$ quark. The current lattice data for the light-sea quark form factors are between our  sea quark results for $u$ and $d$.
\end{abstract}



\footnotetext[0]{\hspace*{-3mm}\raisebox{0.3ex}{$\scriptstyle\copyright$}2013
	Chinese Physical Society and the Institute of High Energy Physics
	of the Chinese Academy of Sciences and the Institute
	of Modern Physics of the Chinese Academy of Sciences and IOP Publishing Ltd}%

\begin{multicols}{2}

\section{Introduction}

Nucleon structure is one of the hottest research objects in hadron physics. A lot of experimental and theoretical efforts have been made on this topic. With the upgrade of the experimental facility, more and more inner information of nucleon
can be detected. By increasing the detecting energy, experimentalists can extract the parton
distribution functions (PDFs) from the deep inelastic scattering as well as the form factors at relatively large momentum transfer
from the elastic scattering. In addition to the valence quark, the information from sea quark contribution to nucleon properties is also very important both analytically and numerically. It is crucial for the leading non-analytic
behavior of the physical quantities.

Strange quark contribution to the nucleon form factors is of special interest because it is purely form sea quark. Lots of experiments from different collaborations in the world have been carried out to measure this quantity as precise as possible
\cite{expriment.SAMPLE.1,expriment.SAMPLE.2,expriment.p2-expri,expriment.review}. There are also many theoretical discussions on the strange form factors \cite{An,Kiswandhi,Riska}. In 2002, we studied the strange form factors with perturbative chiral quark model
(PCQM) and it was found the strange charge form factor is positive, while the strange magnetic form factor is negative \cite{VL}. At that time, the theoretical predictions were quite different because there was no precise experimental measurement yet. 

Theoretically, it is impossible to study nucleon structure using quantum chromodynamics directly because of non-perturbative problem.
Besides the phenomenological quark models, there are two systematic methods in hadron physics. One is the lattice simulation
and the other is effective field theory or chiral perturbation theory. Traditionally,  chiral perturbation theory with dimensional regularization (DR) is valid only at low momentum transfer $Q^2 < 0.1$ GeV$^2$ \cite{theory.chpt-range.1}. It can be applied up to 0.4 GeV$^2$ if vector mesons are included \cite{theory.chpt-range.2}. For the lattice simulation, many quantities simulated on lattice are at large quark (pion) mass. It is necessary to extrapolate lattice data to that at physical pion mass. Instead of DR, we applied the effective field theory with finite-range-regularization (FRR). The vector meson mass, magnetic moments, magnetic form factors, strange form factors, charge radii, first moments of GPDs, nucleon spin, etc can be successfully extrapolated to the physical point \cite{theory.FRR.0,theory.FRR.1,theory.FRR.2,theory.FRR.3,theory.FRR.4,theory.FRR.5,theory.FRR.6,theory.FRR.7,theory.FRR.8,theory.FRR.9}. The obtained strange form factors with FRR are also consistent with our previous results on PCQM \cite{theory.FRR.1,theory.FRR.4}. Recently, the nucleon form factors as well as the sea contribution from light and strange quark were simulated on lattice even at physical pion mass \cite{experiment.lattice.1,s-quark.lattice,experiment.strange.formfactors,Green}. Therefore it is interesting to compare their result with that calculated in the framework of effective field theory.

In these years, we proposed a nonlocal chiral effective Lagrangian which makes it possible to study the hadron properties at relatively large $Q^2$ \cite{Wang1,pingw.quantization,Wang2,fw.nucleon,fw.strange,Salamu.parton}. The nonlocal interaction generates both the regulator which makes the loop integral convergent and the $Q^2$ dependence of form factors at tree level. The obtained electromagnetic form factors and strange form factors of nucleon are very close to the experimental data \cite{fw.nucleon,fw.strange}.  In this paper, we will apply the nonlocal Lagrangian to investigate the light sea quark contribution to the nucleon form factors.
With the quark flow method same as in Ref.~\cite{theory.quarkflow}, we can get the sea and valence quark contribution separately. This method is equivalent to the quenched chiral perturbation theory. In section II, we will introduce the nonlocal
chiral Lagrangian. The sea quark contributions to the nucleon form factors are derived in section III. Numerical results are shown in section IV and finally, section V is a short summary.

\section{Chiral  effective Lagrangian}
The lowest order chiral Lagrangian for baryons, pseudo-scalar mesons and their interaction can be written as \cite{fw.nucleon,Salamu.parton,Jenkins1,Jenkins2},
\begin{equation}
\begin{aligned}
\mathcal{L} ={}& i\, Tr\,\bar{B} \gamma_{\mu}\,\slashed{\mathscr{D}}B -m_B\,Tr\,\bar{B}B \\
{}+{}&\bar{T}_\mu^{abc}(i\gamma^{\mu\nu\alpha}D_\alpha\,-\,m_T\gamma^{\mu\nu})
T_\nu^{abc} \\
{}+{}&\frac{f^2}{4}Tr\,\partial_\mu\Sigma\,\partial^\mu\Sigma^\dagger +D\,Tr\, \bar{B} \gamma_\mu \gamma_5\, \{A_\mu,B\} 
\\ 
{}+{}& F\,Tr\, \bar{B} \gamma_\mu \gamma_5\, [A_\mu,B] \\
{}+{}&\left [\frac{C}{f}\epsilon^{abc}\bar{T}_\mu^{ade}
(g^{\mu\nu}+z\gamma_\mu\gamma_\nu) B_c^e\partial_\nu\phi_b^d+H.C \right ],
\end{aligned}
\end{equation}
where $D$, $F$ and $ C $ are the coupling constants.
The chiral covariant derivative $\mathscr{D}_\mu$ is defined as $\mathscr{D}_\mu
B=\partial_\mu B+[V_\mu,B]$. The pseudo-scalar meson octet
couples to the baryon field through the vector and axial vector
combinations as
\begin{equation}
\begin{aligned}
&V_\mu=\frac12(\zeta\partial_\mu\zeta^\dagger+\zeta^\dagger\partial_\mu\zeta)+\frac{1}{2}ie\mathscr{A}^\mu
(\zeta^\dagger\,Q_c\,\zeta+\zeta\,Q_c\,\zeta^\dagger), \\
&A_\mu=\frac12(\zeta\partial_\mu\zeta^\dagger-\zeta^\dagger\partial_\mu\zeta)-\frac{1}{2}e\mathscr{A}^\mu
(\zeta\,Q_c\,\zeta^\dagger-\zeta^\dagger\,Q_c\,\zeta),
\end{aligned}
\end{equation}
where
\begin{equation}
\zeta^2 = \Sigma =e^{i2\phi/f},\qquad f=93~{\rm MeV}.
\end{equation}
$Q_c$ can be the real charge matrix $\text{diag} (2/3,-1/3,-1/3)$.
$\phi$ and $B$ are the matrices of pseudo-scalar fields and octet baryons.
$\mathscr{A}^\mu$ is the photon field. 
The covariant derivative $D_\mu$ in the decuplet sector is defined as
$D_\nu T_\mu^{abc} = \partial_\nu T_\mu^{abc}+(\Gamma_\nu,T_\mu)^{abc}$,  
where $\Gamma_\nu$ 
is the chiral connection defined as, $(X,T_\mu)=(X)_d^aT_\mu^{dbc}+(X)_d^bT_\mu^{adc}+(X)_d^cT_{\mu}^{abd}$. 
$\gamma^{\mu\nu\alpha}$, $\gamma^{\mu\nu}$ are the antisymmetric matrices expressed as
\begin{equation}
\gamma^{\mu\nu}
=\frac12\left[\gamma^\mu,\gamma^\nu\right]\hspace{.5cm}\text{and}\hspace{.5cm}
\gamma^{\mu\nu\rho}=\frac14\left\{\left[\gamma^\mu,\gamma^\nu\right],
\gamma^\rho\right\}.\,
\end{equation}

The octet, decuplet and octet-decuplet transition magnetic moment
operators are needed in the one loop calculation of nucleon electromagnetic form factors. 
The baryon octet anomalous magnetic Lagrangian is written as
\begin{equation}
\label{eq.octet.anomalous}
\begin{aligned}
\mathcal{L}_{oct}={}&\frac{e}{4m_B}\Big(
c_1\,\Tr\left[\bar{B} \sigma^{\mu\nu}
\left\{F^+_{\mu\nu},B\right\}\right] \\
{}+{}&
c_2\,\Tr\left[\bar{B}\sigma^{\mu\nu} \left[F^+_{\mu\nu},B
\right]\right] \\
{}+{}&c_3\,\Tr\left[\bar{B}\sigma^{\mu\nu}B\right]\,\Tr\left[F^+_{\mu\nu},B\right]
\Big),
\end{aligned}
\end{equation}
where, 
\begin{equation}
F^\dagger_{\mu\nu}=-\frac12\left(\zeta^\dag F_{\mu\nu}\,Q_c\,\zeta+\zeta
F_{\mu\nu}\,Q_c\,\zeta^\dag\right).
\end{equation}
At the lowest order, the contribution of quark $q$ to the nucleon magnetic moments can be obtained by the replacement of the 
charge matrix $Q_c$ with the corresponding diagonal quark matrices $\lambda_q = \text{diag}(\delta_{qu}, \delta_{qd}, \delta_{qs})$.
After the expansion of the above equation, it is found that
\begin{equation}\label{treemag.chpt}
\begin{aligned}
F_2^{p,u}={}&c_1+c_2+c_3,\qquad&
F_2^{n,u}={}&c_3,\\
F_2^{p,d}={}&c_3,\qquad&
F_2^{n,d}={}&c_1+c_2+c_3,\\
F_2^{p,s}={}&c_1-c_2+c_3,\qquad&
F_2^{n,s}={}&c_1-c_2+c_3.\\
\end{aligned}
\end{equation}
Comparing with the results of constituent quark model where
\begin{equation}
\label{treemag.quarkmodel}
\begin{aligned}
F_2^{p,s}={}& 0 ~~\text {and}~~
F_2^{n,s}=0, \\
\end{aligned}
\end{equation}
we can get
\begin{equation}
\mathrm{c_3}=\mathrm{c_2}-\mathrm{c_1}.
\end{equation}
The transition magnetic operator is
\begin{multline}
\mathcal{L}_{tr}=i\frac{e}{4m_N}\mu_T F_{\mu\nu}
\Big(
\epsilon^{ijk} Q_c^{il}\bar{B}^{jm}\gamma^\mu\gamma_5\,
T^{\nu,klm} \\
+
\epsilon^{ijk} Q_c^{li}\bar{T}^{\mu,klm}\,\gamma^\nu\gamma_5\,B^{mj}
\Big).
\end{multline}
The effective decuplet anomalous magnetic moment operator can be expressed as
\begin{equation}\label{eq:ci}
\mathcal{L}_{dec}=-\frac{ieF_2^T}{4M_T}\bar{T}_\mu^{abc}\sigma^{\rho\lambda}F_{\rho\lambda}T^{\mu,abc}.
\end{equation}
The anomalous magnetic moments of baryons can also be expressed in terms of quark magnetic moments $\mu_q$.
For example, $\mu_p = \frac43 \mu_u - \frac13 \mu_d$, $\mu_n = \frac43 \mu_d - \frac13 \mu_u$, 
$\mu_{\Delta^{++}} = 3\mu_u$. Using the SU(3)
symmetry $\mu_u=-2\mu_d = -2\mu_s$, $\mu_T$ and $F_2^T$ as well as $\mu_q$ can be written in terms of $c_1$
or $c_2$. For example, $\mu_u = \frac23 c_1$, $\mu_T= 4 c_1$, $F_2^{\Delta^{++}} = \mu_{\Delta^{++}}-2 = 2c_1-2$.

The gauge invariant non-local Lagrangian can be obtained using the method in \cite{pingw.quantization,fw.nucleon,fw.strange}. 
For instance, the local interaction including $\pi$ meson can be written as
\begin{equation}
\mathcal{L}_{\pi}^{local}=\int\!\,dx \frac{D+F}{\sqrt{2}f} \bar{p}(x)\gamma^\mu\gamma_5\,n(x)(\partial_\mu+ie\,\mathscr{A}_\mu(x)) \pi^+(x).
\end{equation}
The corresponding nonlocal Lagrangian is expressed as
\begin{align}\label{eq:nonlocal}
{\cal L_{\pi}}^{nl}&=\int\!\,dx\int\!\,dy\frac{D+F}{\sqrt{2}f}\bar{p}(x)\gamma^\mu\gamma_5 n(x)F(x-y) \\
&\times \mathrm{exp}[ie\int_x^y dz_\nu\int\!\,da\,\mathscr{A}^\nu(z-a)F(a)] \nonumber \\
&\times (\partial_\mu\,+ie\int\!\,da\,\mathscr{A}_\mu(y-a))F(a)\big)\pi^+(y),
\end{align}
where $F(x)$ is the correlation function.
To guarantee the gauge invariance, the gauge link is introduced in the above Lagrangian.
The regulator can be generated automatically with correlation function.

With the same idea, the nonlocal electromagnetic interaction can also be obtained.
For example, the local interaction between proton and photon is written as
\begin{multline}
{\cal L}_{EM}^{local} =  -e \bar{p}(x) \gamma^\mu p(x) \mathscr{A}_\mu(x) s \\
+\frac{(c_1-1)e}{4m_N} \bar{p}(x)\sigma^{\mu\nu}p(x)F_{\mu\nu}(x).
\end{multline}
The corresponding nonlocal Lagrangian is expressed as
\begin{multline}
 {\cal L}_{EM}^{nl} = -e \int da \bar{p}(x) \gamma^\mu p(x) \mathscr{A}_\mu(x-a)F_1(a) \\
+ \frac{(c_1-1)e}{4m_N}\int da \bar{p}(x)\sigma^{\mu\nu}p(x)F_{\mu\nu}(x-a)F_2(a),
\end{multline} 
where $F_1(a)$ and $F_2(a)$ are the correlation functions for the nonlocal electric and magnetic interactions.

The form factors at tree level which are momentum dependent can be easily obtained with the Fourier transformation
of the correlation function. As in our previous work \cite{fw.nucleon,fw.strange}, the correlation function is chosen so that the charge and magnetic form factors
at tree level have the same the momentum dependence as nucleon-pion vertex, i.e. $G_M^{\rm tree}(q)=c_1G_E^{\rm tree}(q) = c_1\tilde{F}(q)$,
where $\tilde{F}(q)$ is the Fourier transformation of the correlation function $F(a)$.
The corresponding function of $\tilde{F}_1(q)$ and $\tilde{F}_2(q)$ is then expressed as
\begin{equation}
\tilde{F}_1^p(q)=\tilde{F}(q)\frac{4m_N^2+c_1Q^2}{4m_N^2+Q^2},~~~\tilde{F}_2^p(q)=\tilde{F}(q)\frac{4m_N^2}{4m_N^2+Q^2}, 
\end{equation}
where $Q^2 = -q^2$ is the momentum transfer.
From Eq.~(\ref{eq:nonlocal}), two kinds of couplings
between hadrons and one photon can be obtained. One is the normal one expressed as
\begin{multline}
{\cal L}^{nor}=ie\int\!\,dx\int\!\,dy\frac{D+F}{\sqrt{2}f}\bar{p}(x)\gamma^\mu\gamma_5 n(x)F(x-y)\pi^+(y) \\
\times \int\!\,da\,\mathscr{A}_\mu(y-a)F(a),
\end{multline}
This interaction is similar as the traditional local Lagrangian except the correlation function. 
The other one is the additional interaction obtained by the expansion of the gauge link, expressed as
\begin{multline}
{\cal L}^{add}=ie\int\!\,dx\int\!\,dy\frac{D+F}{\sqrt{2}f}\bar{p}(x)\gamma^\mu\gamma_5 n(x)F(x-y) \\
\times \int_x^y dz_\nu\int\!\,da\,\mathscr{A}^\nu(z-a)F(a)\partial_\mu \pi^+(y) .                    
\end{multline}

\section{Electromagnetic form-factors}

The contribution from the quark flavor $f$ $(f=u,d,s)$ to the Dirac and Pauli form factors of nucleon are defined as
\begin{multline}
\label{eq:f1f2}
<N(p^\prime)|J_\mu^f|N(p)>=\bar{u}(p^\prime)\\
\left\{
\gamma^\mu F_1^{f}(Q^2)+
\frac{i\sigma^{\mu\nu} q_\nu}{2m_N}F_2^{f}(Q^2)
\right\}u(p),
\end{multline}
where $q=p^\prime-p$.
The electromagnetic form-factors are defined as the combinations of the above ones for each flavor as
\begin{equation}
\begin{aligned}
G_E^{f}(Q^2) &{}=F_1^{f}(Q^2)-\frac{Q^2}{4m_N^2}F_2^{f}(Q^2),\\
G_M^{f}(Q^2) &{}=F_1^{f}(Q^2) +F_2^{f}(Q^2).
\end{aligned}
\end{equation}
In this manuscript, we will investigate the sea quark contribution to the nucleon electromagnetic form
factors from $u$, $d$ and $s$.
According to the Lagrangian, the one loop Feynman diagrams which contribute to the nucleon electromagnetic 
form factors are plotted in Fig.~\ref{fig.loop.diagrams}. 
\end{multicols}

\vspace{4mm}
\ruleup
\begin{center}
		\includegraphics[width=0.8\textwidth]{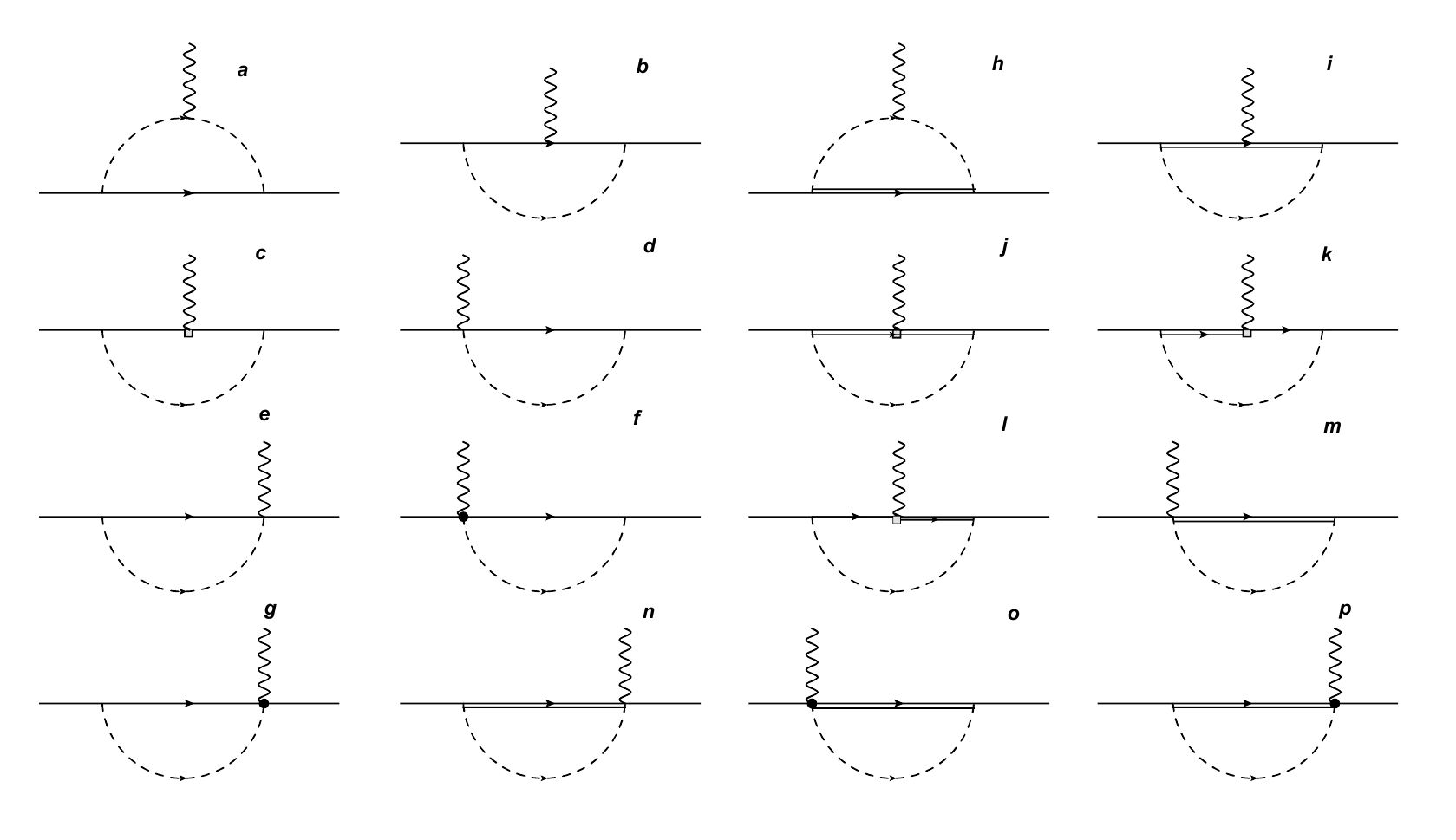}
		\figcaption{\label{fig.loop.diagrams} 
		One-loop Feynman diagrams for the nucleon electromagnetic form factors. The solid, double-solid, dashed and wave lines 
		are for the octet baryons, decuplet baryons, pseudo-scalar mesons and photons, respectively. The rectangle and black dot represent 
		magnetic and additional interacting vertex.}
\end{center}
\ruledown
\vspace{4mm}

\begin{multicols}{2}

From the Lagrangian, the coupling constants between baryons and mesons (coefficients) in Fig.~1 can be obtained.
For each diagram in Fig.~1, there exist quenched and disconnected diagrams.  	
In order to obtain the pure sea quark contribution, we need to get the coefficients for the disconnected diagrams.
The coefficients for the quenched and disconnected loop diagrams can be got separately as 
in Ref.~\cite{theory.quarkflow} using the quark flows of Fig.~\ref{fig.quarkflow}. The obtained coefficients are the same as those 
extracted within the graded symmetry formalism in quenched chiral perturbation theory \cite{theory.formal-quench}. 
In Fig.~\ref{fig.quarkflow}, we plot the diagram for the $\pi^+$ rainbow diagram using quark flows as an example 
to show the method of separating the quenched and sea quark contribution.
The coefficients for the $\pi^+$ loop diagram in full QCD is $(D+F)^2$. The coefficient of Fig.~\ref{fig.quarkflow}b
for the sea quark contribution is the same as that of Fig.~\ref{fig.quarkflow}c for the $K^+$ loop.
The coefficient for quenched sector can be obtained by subtracting the coefficient of sea diagram from the total one. 
The coefficients of $u$, $d$ and $s$ quark for both quenched and sea quark flow diagram are listed 
in Table.~\ref{chpt.coefficients.f1}. For $\pi^0$ case, the first and second rows are for $u\bar{u}$ and $d\bar{d}$, respectively

\begin{center}
\includegraphics[width=0.45\textwidth]{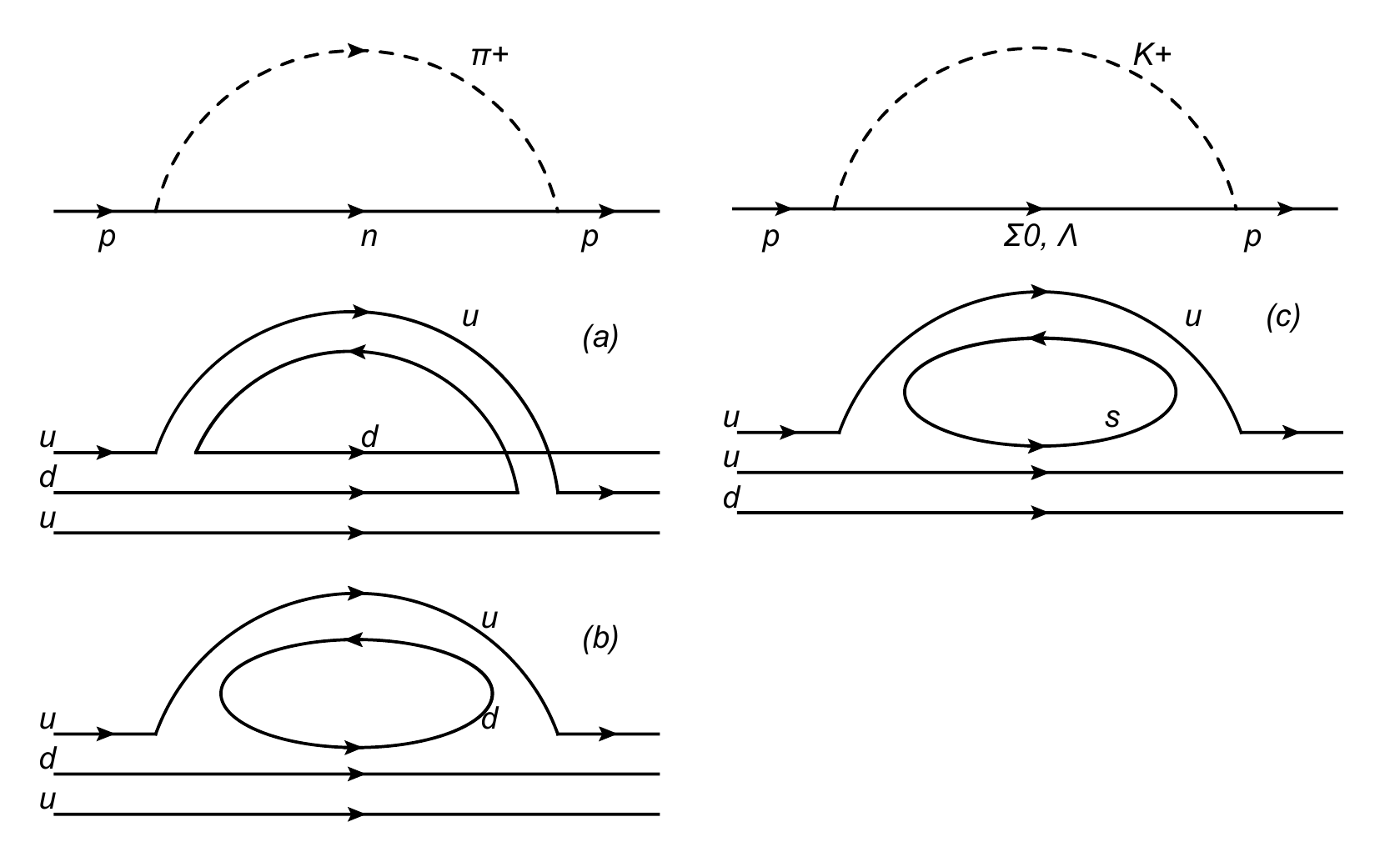}
\figcaption{\label{fig.quarkflow} 
	Quark flow diagrams for $\pi^+$ and $K^+$. (a) is the quenched diagram. (b) and (c) are the disconnected 
	sea diagrams for $\pi^+$ and $K$, respectively.}
\end{center}

With the Lagrangian we can get the matrix element of Eq.~(\ref{eq:f1f2}).
In this section, we will only show the expressions for the intermediate octet baryon part. 
For the intermediate decuplet baryon, the expressions are similar but more complicated. 

\begin{center}
\tabcaption{ \label{chpt.coefficients.f1}  The coefficients for the quenched and sea diagrams as well as the total one for Fig.~2.}
\footnotesize	
	\begin{tabular*}{80mm}{c@{\extracolsep{\fill}}ccc}
		\hline
		meson & full QCD & quenched diagram & sea diagram \\ \hline
		$\pi^0$ & $\frac{1}{2} (D+F)^2$ & $-\frac{D^2}{3}+2 D F-F^2$ & \begin{tabular}[c]{@{}c@{}}$\frac{D^2}{3}+F^2$\\ $\frac{1}{2} (D-F)^2$\end{tabular} \\ \hline
		$\pi^+$ & $(D+F)^2$ & $\frac{1}{3} \left(D^2+6 D F-3 F^2\right)$ & $\frac{2}{3} \left(D^2+3 F^2\right)$ \\ \hline
		$\pi^-$ & $0$ & $-(D-F)^2$ & $(D-F)^2$ \\ \hline
		$K^0$ & $(D-F)^2$ & $0$ & $(D-F)^2$ \\ \hline
		$K^+$ & $\frac{2}{3} \left(D^2+3 F^2\right)$ & $0$ & $\frac{2}{3} \left(D^2+3 F^2\right)$ \\ \hline
	\end{tabular*}
\vspace{0mm}
\end{center}
\vspace{0mm}

The sea quark contributions of Fig.~1a for quark $u$ $d$ and $s$ are written as
\begin{align} 
\Gamma_a^u={}&
(\frac{4 D^2}{3}-2 D F+2 F^2) I_a^{N\pi},\\
\Gamma_a^d={}&
(\frac{7 D^2}{6}-D F+\frac{5 F^2}{2})I_a^{N\pi},
\\
\Gamma_a^s={}&
\frac{1}{6} (D+3 F)^2 I_a^{\Lambda K}+
\frac{3}{2} (D-F)^2 I_a^{\Sigma K},
\end{align} 
where the integral $I_a^{BM}$ is expressed as
\begin{multline} 
I_a^{BM}=\frac{1}{2f^2}\bar{u}(p^\prime)\tilde F(q)\int\!\frac{d^4 k}{(2\pi)^4}(\slashed{k}+\slashed{q})\gamma_5\,\tilde F(q+k)
\frac{1}{D_K(k+q)}\\
(2k+q)^\mu\frac{1}{D_M(k)}\frac{1}{\slashed{p}-\slashed{k}-m_B}
(-\slashed{k}\gamma_5)\tilde F(k)u(p).
\end{multline}
$D_M(k)$ is defined as 
\begin{equation} 
D_M(k)=k^2-m_M^2+i\epsilon.
\end{equation}
$m_B$ and $m_M$ are the masses for the intermediate $B$ baryon and $M$ meson, respectively. 
The contributions of Fig.~1b are expressed as
\begin{align} \nonumber
\Gamma_b^u={}&
\frac{2 i m_B} {9 \left(4 m_B^2+Q^2\right)} [c_1 \left(-11 D^2+24 D F-9 F^2\right) \\
& +6 c_2 \left(2 D^2-3 D F+3 F^2\right)]
I_b^{N \pi}, \\ \nonumber
\Gamma_b^d={}&
-\frac{i m_B}{9 \left(4 m_B^2+Q^2\right)}  [c_1 \left(17 D^2-42 D F+9 F^2\right)\\
& -3 c_2 \left(7 D^2-6 D F+15 F^2\right) ]
I_b^{N \pi}, \\ \notag
\Gamma_b^s={}&
\frac{i \left(c_1+3 c_2\right) m_B (D+3 F)^2}{9 \left(4 m_B^2+Q^2\right)}
I_b^{\Lambda \Lambda K} \\ 
& + \frac{3 i \left(c_2-c_1\right) m_B (D-F)^2}{4 m_B^2+Q^2}
I_b^{\Sigma K}, 
\end{align}
where the integral $I_b^{BM}$ is written as
\begin{multline} 
I_b^{BM}=\frac{1}{2f^2}\bar{u}(p^\prime)\tilde F(q)\int\!\frac{d^4 k}{(2\pi)^4}\slashed{k}\gamma_5\,\tilde F(k)\frac{1}{D_M(k)}
\\ \frac{1}{\slashed{p'}-\slashed{k}-m_B}\gamma^\mu 
\frac{1}{\slashed{p}-\slashed{k}
-m_B}\frac{\slashed{k}\gamma_5}{\sqrt{2}f}\tilde F(k)u(p).
\end{multline}
Fig.1c is similar as Fig.1b except for the magnetic interaction. The contributions of this diagram are written as
\begin{align} 
\Gamma_c^u={}&
\bigg(\frac{2 i m_B \left(c_1 \left(-11 D^2+24 D F-9 F^2\right)\right)}{9 \left(4 m_B^2+Q^2\right)} \notag \\
{}+{}&\frac{2 i m_B \left(6 c_2 \left(2 D^2-3 D F+3 F^2\right)\right)}{9 \left(4 m_B^2+Q^2\right)}
\bigg)
I_c^{N \pi},\\ \notag
\Gamma_c^d={}&
\bigg(-\frac{i m_B \left(c_1 \left(17 D^2-42 D F+9 F^2\right)\right)}{9 \left(4 m_B^2+Q^2\right)} \\
{}+{}&\frac{i m_B \left(-3 c_2 \left(7 D^2-6 D F+15 F^2\right)\right)}{9 \left(4 m_B^2+Q^2\right)}
\bigg)
I_c^{N \pi}, \\ 
\Gamma_c^s={}&
\frac{i \left(c_1+3 c_2\right) m_B (D+3 F)^2}{9 \left(4 m_B^2+Q^2\right)}
I_c^{\Lambda \Lambda K} \nonumber \\  
{}+{}&\frac{3 i c_3 m_B (D-F)^2}{4 m_B^2+Q^2}
I_c^{\Sigma K}, 
\end{align} 
where $I_c^{\Lambda K}$ is expressed as
\begin{multline}
I_c^{BM}=\frac{1}{2f^2}\bar{u}(p^\prime)\tilde F(q)\int\!\frac{d^4 k}{(2\pi)^4}\slashed{k}\gamma_5\tilde F(k)
\frac{1}{\slashed{p'}-\slashed{k}-m_B} \\
\frac{\sigma^{\mu\nu}q_{\nu}}{2m_\Lambda}
\frac{1}{\slashed{p}-\slashed{k}-m_B}\frac{i}{D_M(k)}
\slashed{k}\gamma_5\tilde F(k)u(p).
\end{multline}
Fig.~1d and 1e are the Kroll-Ruderman diagrams. The contributions from these two diagrams are written as 
\begin{align} 
\Gamma_{d+e}^u={}&
\left(-\frac{4 D^2}{3}+2 D F-2 F^2\right)
I_{d+e}^{N \pi}, \\
\Gamma_{d+e}^d={}&
(-\frac{7 D^2}{6}+D F-\frac{5 F^2}{2})
I_{d+e}^{N \pi}, \\
%
\Gamma_{d+e}^s={}&
-\frac{1}{6} (D+3 F)^2
I_{d+e}^{\Lambda K}
-\frac{3}{2} (D-F)^2
I_{d+e}^{\Sigma K},
\end{align}
where
\begin{equation}
\begin{aligned}
I_{(d+e)}^{BM}={}&\frac{1}{2f^2}\bar{u}(p^\prime)\tilde F(q)\int\!\frac{d^4 k}{(2\pi)^4}\slashed{k}\gamma_5\tilde F(k)
\frac{1}{\slashed{p'}-\slashed{k}-m_B} \\
& \frac{1}{D_M(k)}\gamma^\mu\gamma_5 \tilde F(k)u(p) \\
{}+{}&\frac{1}{2f^2}\bar{u}(p^\prime)\tilde F(q)\int\!\frac{d^4 k}{(2\pi)^4}\gamma^\mu\gamma_5\tilde F(k)\frac{1}{\slashed{p}-\slashed{k}-m_B} \\
& \frac{1}{D_M(k)}\slashed{k}\gamma_5\tilde F(k)u(p).
\end{aligned}
\end{equation} 
Fig.~1f and 1g are the additional diagrams which are generated from the expansion of the gauge link terms.
The contributions of these two additional diagrams with intermediate octet hyperons are expressed as
\begin{align} 
\Gamma_{f+g}^u={}&
(-\frac{4 D^2}{3}+2 D F-2 F^2)
I_{f+g}^{N \pi}, \\
\Gamma_{f+g}^d={}&
(-\frac{7 D^2}{6}+D F-\frac{5 F^2}{2})
I_{f+g}^{N \pi} ,\\
\Gamma_{f+g}^s={}&
-\frac{1}{6} (D+3 F)^2
I_{f+g}^{\Lambda K}
-\frac{3}{2} (D-F)^2
I_{f+g}^{\Sigma K},
\end{align}
 where
\begin{equation}
\begin{aligned}
I_{f+g}^{BM}={}&
\frac{1}{2f^2}\,\bar{u}(p^\prime)\tilde F(q)\int\!\frac{d^4 k}{(2\pi)^4}\slashed{k}\gamma_5\tilde F(k)\frac{1}{\slashed{p'}
-\slashed{k}-m_B} \\
& \frac{1}{D_M(k)}(-\slashed{k}+\slashed{q})\gamma_5\frac{(2k-q)^\mu}{2kq-q^2}
[\tilde F(k-q)-\tilde F(k)]u(p) \\
{}+{}&
\frac{1}{2f^2}\bar{u}(p^\prime)\tilde F(q)\int\!\frac{d^4 k}{(2\pi)^4}(\slashed{k}+\slashed{q})\gamma_5\frac{(2k+q)^\mu}{2kq+q^2} \\
&[\tilde F(k+q)-\tilde F(k)] \frac{1}{\slashed{p}-\slashed{k}-m_B}
 \frac{1}{D_M(k)}\slashed{k}\gamma_5F(k)u(p).
\end{aligned}
\end{equation}
Using Package-X \cite{tool.packagex} to simplify the $\gamma$ matrix algebra, we can get the separate expressions for the Dirac and Pauli form factors. In the next section, we will discuss numerical results.

\begin{center}
	\includegraphics[width=.44\textwidth]{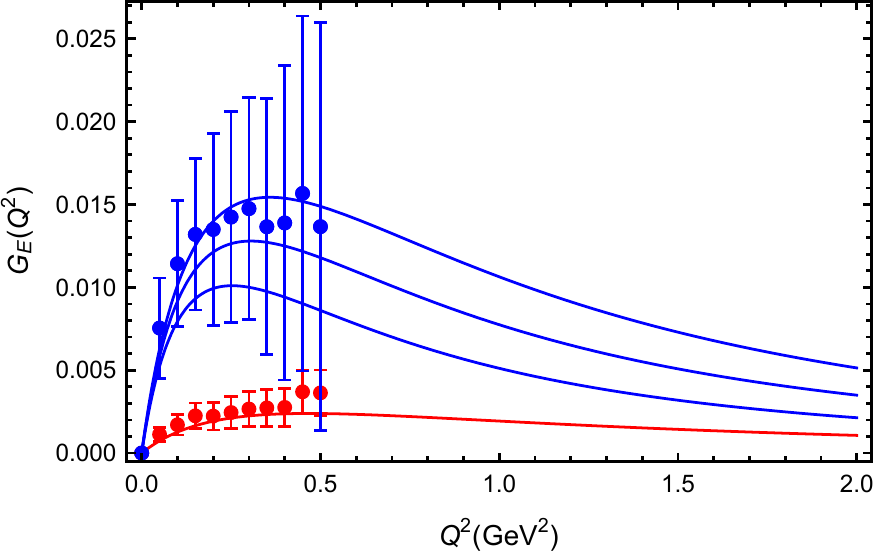}
	\figcaption{\label{fig.sea.ge.us} The sea quark contributions of $u$ to the proton electric form factor versus momentum transfer $Q^2$.
	Three blue lines from up to bottom are for $\Lambda=$ 1.0, 0.9 and 0.8 \gev, respectively.		
	The red line is for the strange quark result with $\Lambda =0.9$ GeV. 
	The data with error bars are from lattice simulation \cite{s-quark.lattice}.
	}
\end{center}

\section{Numerical Results}

In the numerical calculations, the parameters are chosen as $D=0.76$ and $F=0.50$ ($g_A=D+F=1.26$).
The coupling constant ${C}$ is chosen to be $1$ which is the same as in Ref.~\cite{c-get-1,fw.nucleon}.
The off-shell parameter $z$ is chosen to be $z=-1$ \cite{nath.decuplet}.
The low energy constants $c_1$ and $c_2$ are determined to be $2.057$ and $0.748$, which give the experimental moments of 
$\mu_p = 2.793$ and $\mu_n = -1.913$.
The covariant regulator is chosen to be dipole form \cite{fw.nucleon,fw.strange,Salamu.parton}
\begin{equation}
\tilde{F}[k]=\frac{\Lambda^4}{(\Lambda^2+m_j^2-k^2)^2},
\end{equation} 
where $m_j$ is the meson mass for the baryon-meson interaction and it is zero for the hadron-photon interaction.
It was found that when $\Lambda$ was around 0.90 \gev, the results are very close to the experimental nucleon form factors.

In Fig.~3, we plot the sea contribution of $u$ quark with unity charge to the proton electric form factor. Three blue 
lines from up to bottom are for $\Lambda=1.0,~0.9$ and $0.8~\mathrm{GeV}$, respectively. As a comparison, 
the central result for the strange quark is also plotted 
in the figure with red line. The solid dots with error bars are lattice data from Ref.~\cite{s-quark.lattice}.
Since we did not include the valence contribution of $u$ quark in proton, the electric form factor of $u$ quark is 
zero when $Q^2=0$. It then increases with the increasing $Q^2$.
When $Q^2$ is larger than about 0.3 $\mathrm{GeV}^2$, it deceases with $Q^2$. From the figure, 
one can see the strange form factor can be described very well. The $u$ quark result
is a little smaller than the lattice data. We should mention that lattice data are for the light quark and it was
assumed that $u$ and $d$ had the same sea contribution. Therefore, lattice data for the light quark can be approximately
treated as an average of the $u$ and $d$ contribution. 

\begin{center}
\includegraphics[width=.44\textwidth]{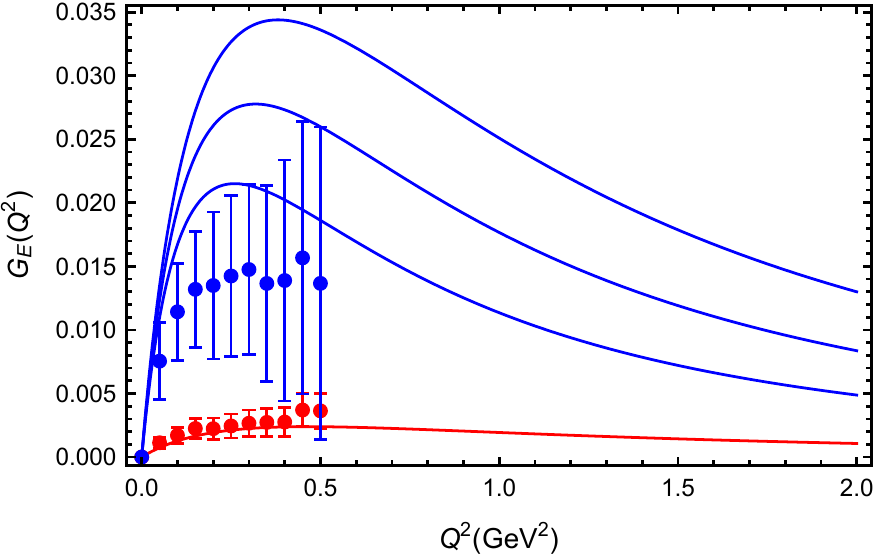}
\figcaption{ \label{fig.sea.ge.ds}
	The sea quark contributions of $d$ to the proton electric form factor versus momentum transfer $Q^2$.
Three blue lines from up to bottom are for $\Lambda=$ 1.0, 0.9 and 0.8 \gev, respectively.		
The red line is for the strange quark result with $\Lambda =0.9$ GeV. 
The data with error bars are from lattice simulation \cite{s-quark.lattice}.}
\end{center}

\begin{center}
\includegraphics[width=.44\textwidth]{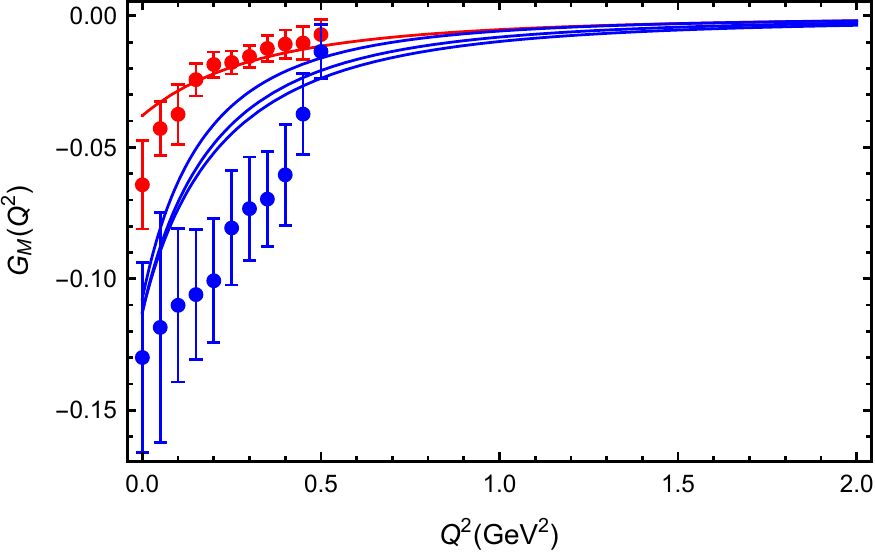}
\figcaption{\label{fig.sea.gm.us}
	Same as Fig.3 but for magnetic form factor.}
\end{center}

The sea contribution to the proton electric form factor from $d$ is plotted in Fig.~4. Similar as in the $u$
quark case, the electric form factor first increases from 0 and then decreases with the increasing $Q^2$.
It can be seen clearly that 
our calculated sea quark contribution is obviously larger than the lattice data. The larger sea contribution from $d$ 
quark than that from $u$ quark is due to the fact that there is no intermediate octet contribution for $u$ quark.
The only contribution for $u$ quark is from the decuplet intermediate states. Similar results can be found
for the $\bar{d}-\bar{u}$ asymmetry in proton, where $\bar{d}$ is excess $\bar{u}$ \cite{Salamu.asy.1,Salamu.asy.2,Salamu.asy.3,Salamu.asy.4}. Though there is obviously difference
of the sea quark contribution between $u$ and $d$, both of them are much larger than that of strange quark contribution.
The strange electric form factor is about 5-10 times smaller due to the suppression of the $K$ meson loop.

The sea contribution to the proton magnetic form factor from $u$ and $d$ quark with unit charge are 
plotted in Fig.~5 and Fig.~6.
Again, the calculated strange magnetic form factor is in good agreement with the lattice data. All the quark magnetic 
form factors increase monotonously with the increasing $Q^2$. For the $u$ quark contribution,
the absolute values are smaller than the light quark result of lattice simulation, while for the $d$ quark contribution, 
the absolute values are larger than lattice data. The absolute contributions of both $u$ and $d$ quark are larger than
that of strange quark, especially at small $Q^2$. At $Q^2=0$, the magnetic moments of the sea quark $u$
and $d$ are $-0.11$ and $-0.39$, respectively, while the strange magnetic moment is about $-0.04$.

\begin{center}
\includegraphics[width=.44\textwidth]{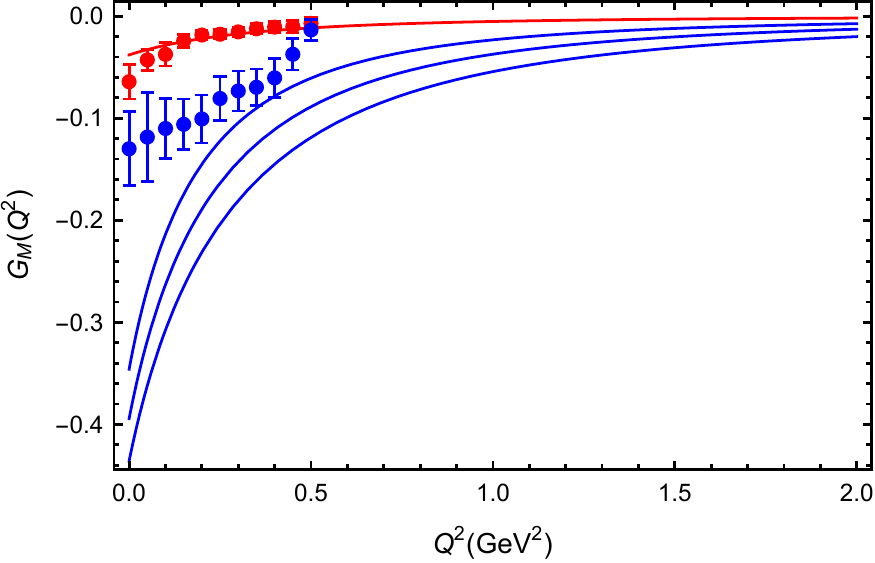}
\figcaption{\label{fig.sea.gm.ds}
	Same as Fig.~4 but for magnetic form factor.}
\end{center}

From the above figures, one can see that the sea quark contributions of $u$ and $d$ quark are quite different.
In both electric and magnetic form factor cases, the absolute value of sea quark contribution of $d$ is much larger
than that of $u$. This is because for proton, there are two up quarks and one down quark. The $u$ quark in the
loop diagram can only form a decuplet state and there is no intermediate octet contribution to the sea quark form
factors of $u$. We should mention this difference between light-sea quark form factors is not due to 
the mass difference of $u$ and $d$ quark.
In fact, in our calculation the masses of $\pi^0$, $\pi^+$ and $\pi^-$ are degenerate. 
It is straightforward that the sea quark form factor
of $u$ ($d$) in proton is the same as that of $d$ ($u$) in neutron if the masses of proton and neutron are the same.
The mass difference between proton and neutron will lead to a small charge symmetry violation, i.e. 
a small difference between $G^u_p$ and $G^d_n$ ($G^d_p$ and $G^u_n$). The large difference between 
$G^u_p$ and $G^d_p$ is because of the effect of non-perturbative valence quark environment
instead of mass difference between $u$ and $d$.

If the mass of the three $u$ quark states is taken to be degenerate with nucleon mass,
and $\eta$ mass taken to be degenerate with $\pi$ mass. the sea contribution from
$u$ and $d$ quark in proton will be the same. This is the artifact of the current lattice simulation.
Physically, three $u$ quark can not form a octet baryon and the mass of $\eta$ is much heavier than
that of $\pi$. Therefore, it is very interesting and challenging to get the flavor asymmetry 
from the lattice with the full-QCD simulation.

\section{Summary}
In this work, we applied the nonlocal chiral effective Lagrangian to study the sea quark contribution of
light quark to the proton electromagnetic form factors. Since the signs of the sea quark form factors
are the same for $u$, $d$ and $s$ quark, this calculation is helpful to understand the experimental values of strange
form factors. It is also interesting to compare our result with that from the lattice simulation.
In our calculation, the parameter $\Lambda$ in the regulator is the same as the previous one which is 
determined by fitting the nucleon form factors.
The low energy constants $c_1$ and $c_2$ are determined by the experimental magnetic moments of proton and neutron.
Therefore, in calculating the sea quark form factors, there is no free parameters to be adjusted. 
Our results show the electric form factor of light sea quark with unity charge is positive, while the magnetic form factor
is negative. Compared with the strange form factors, the absolute value of the light quark form factors
are much larger. For both electric and magnetic form factors, the contribution from $d$ quark is 
larger than that from $u$ quark. The current lattice data for the light-sea quark form factors are
between our results for $u$ and $d$. Therefore, it is interesting if this flavor asymmetry can
be obtained from lattice with full-QCD simulation.

\end{multicols}

\vspace{-1mm}
\centerline{\rule{80mm}{0.1pt}}
\vspace{2mm}

\begin{multicols}{2}


\end{multicols}
	
\end{document}